\documentclass{article}

\usepackage{arxiv}

\usepackage[utf8]{inputenc} 
\usepackage[T1]{fontenc}    
\usepackage{hyperref}       
\usepackage{url}            
\usepackage{booktabs}       
\usepackage{amsfonts}       
\usepackage{nicefrac}       
\usepackage{microtype}      
\usepackage{csquotes}
\usepackage{graphicx}
\usepackage{doi}
\usepackage{amssymb}
\usepackage{multirow}
\usepackage{tabularx}
\usepackage{longtable}
\usepackage{graphicx}
\usepackage{times}
\PassOptionsToPackage{hyphens}{url}
\usepackage[usenames]{color}
\usepackage[dvipsnames]{xcolor}
\usepackage{float}
\usepackage{textcomp}
\usepackage{units}

\usepackage{enumerate} 
\usepackage{wrapfig}
\usepackage{subfigure}
\usepackage{fancyhdr}
\usepackage{afterpage}
\usepackage{caption}
\usepackage{multicol}
\usepackage{hyperref}
\usepackage[british,UKenglish,USenglish]{babel}
\usepackage{gensymb}
\usepackage[none]{hyphenat}
\usepackage {lineno}
\usepackage{eurosym}
\usepackage{amsmath}
\makeatletter
\renewcommand\thefootnote{\textsuperscript{\@fnsymbol\c@footnote}}
\let\old@thanks\thanks 
\DeclareRobustCommand\thanks[2][]{
  \AddToHook{begindocument/end}{
    \if\relax#1\relax%
      \footnotemark%
    \else%
      \protect\refstepcounter{footnote}\protect\label{#1}%
    \fi%
    \protected@xdef\@thanks{%
      \@thanks\protect\footnotetext[\the\c@footnote]{#2}%
    }%
  }%
}
\AddToHook{begindocument/begin}{\let\thanks\old@thanks} 

\let\old@maketitle\maketitle
\def\maketitle{\old@maketitle\def\thefootnote{\@arabic\c@footnote}}
\makeatother

\title{Physical foundations for trustworthy medical imaging: a review for artificial intelligence researchers}

\thanks[1]{Advanced Computing and e-Science Group, Institute of Physics of Cantabria (IFCA), CSIC - UC, Spain.}

\thanks[2]{AI Technology for Life, Department of Information and Computing Sciences, Department of Biology, Utrecht University, Utrecht, Netherlands.}

\thanks[3]{Department of Radiology, The Netherlands Cancer Institute, Amsterdam, Netherlands.}

\thanks[4]{Department of Radiology, León University Health Care Complex, León, Spain.}
\thanks[5]{Department of Morphology and Cell Biology and Group of Peripheral Nervous System and Sensory Organs, University of Oviedo, Oviedo, Spain.}
\newcommand*\samethanks[1][\value{footnote}]{\footnotemark[#1]}
\author{Miriam Cobo\ref{1}\ref{2}\ref{3} \\
\texttt{cobocano@ifca.unican.es} \\
    \And
	David Corral Fontecha\ref{4}\ref{5}
    \And
	Wilson Silva\ref{2}\ref{3}\thanks{Both authors share Senior authorship.}
	\And
	Lara Lloret Iglesias\ref{1}\samethanks\\
}

\renewcommand{\headeright}{}
\renewcommand{\undertitle}{}
\renewcommand{\shorttitle}{}

\hypersetup{
pdfauthor={Miriam Cobo, Lara Lloret Iglesias},
}
\usepackage{indentfirst}
\usepackage{xcolor}
\DeclareUnicodeCharacter{2212}{-}

\usepackage[usestackEOL]{stackengine}
\usepackage{mathtools}
\usepackage{todonotes}
\begin{document}
\maketitle

\begin{abstract}

Artificial intelligence in medical imaging has seen unprecedented growth in the last years, due to rapid advances in deep learning and computing resources. Applications cover the full range of existing medical imaging modalities, with unique characteristics driven by the physics of each technique. Yet, artificial intelligence professionals entering the field, and even experienced developers, often lack a comprehensive understanding of the physical principles underlying medical image acquisition, which hinders their ability to fully leverage its potential. The integration of physics knowledge into artificial intelligence algorithms enhances their trustworthiness and robustness in medical imaging, especially in scenarios with limited data availability. In this work, we review the fundamentals of physics in medical images and their impact on the latest advances in artificial intelligence, particularly, in generative models and reconstruction algorithms. Finally, we explore the integration of physics knowledge into physics-inspired machine learning models, which leverage physics-based constraints to enhance the learning of medical imaging features.

\end{abstract}


\section{Introduction}

Since the accidental discovery of X-rays by  Wilhelm Roentgen in 1895 \cite{samei2022medical}, medical imaging has evolved to play an essential role in daily healthcare. Applications expand beyond diagnostic purposes to treatment planning, monitoring the progression of diseases, real-time guided interventions, together with research and educational purposes, such as functional imaging, and construction of population atlases.

Crucial aspects of medical images include finding a compromise between patient safety and image quality in the acquisition power and energy levels, together with the amount of time required to generate the images, since patient motion hinders the acquisition of higher resolution medical images. Yet, in some modalities such as nuclear medicine, the acquisition time is directly limited by physics \cite{bushberg2021essential}. These factors are taken into account to find an optimal balance in the clinics, and they are essential in medical image reconstruction. Hence, medical images are optimized to perform the required task at the necessary image quality that ensures accurate diagnosis \cite{bushberg2021essential}.

Health systems are transitioning to personalized, preventive, predictive, participative and precision (5P) medicine \cite{blobel2019does}. In medical imaging, artificial intelligence (AI) is playing a key role in transforming diagnostics and therapeutics to enable precision medicine. Yet, AI applications in medical imaging are constrained by the scarcity of high-quality, large-scale labeled data. Synthetic data offers a potential approach to mitigate biases and improve algorithmic fairness and reliability \cite{van2024can}, but also poses technical and ethical challenges \cite{koetzier2024generating}. AI-generated synthetic data aims to resemble real data, and to do so AI generative models are required to capture the complexity of medical images, which is given by their physical properties. 

There is a clear gap between research in AI for medical imaging applications and clinical translation, which is hindered by challenges in generalization, standardization and privacy limitations \cite{cobo2023enhancing}. Moreover, discrepancies in how neural networks learn from different domains, in particular the adaptation between medical imaging and natural images domain are often overlooked when developing AI algorithms \cite{konz2024effect}. Current research efforts are moving towards vision foundation models for medical imaging, but many challenges remain before clinical translation \cite{azad2023foundational}. 
A universal foundation model approach would struggle to achieve state-of-the-art performance in many medical image analysis tasks, because of the significant variability in organ and structural features, such as texture, shape, size and topology, in addition to variations in the physical properties across imaging modalities \cite{zhang2024challenges}. Large language models (LLMs) are gaining attention in radiology to automate structure reporting, demonstrating promising results \cite{busch2024large}. The combination of LLMs with vision foundation models holds great potential for health applications \cite{zhang2024challenges}. However, widespread adoption of such algorithms in radiology is subject to compliance with regulatory frameworks, which will determine the degree to which AI-powered technologies are adopted in the near future in the clinical practice.
Any AI algorithm needs to be trustworthy to ensure its safe use in the clinics, which involves a wide range of factors including robustness, security, transparency, explainability, fairness, and safety \cite{ali2023explainable, li2023trustworthy}. 

The 2024 Nobel Prize in Physics, awarded to John J. Hopfield and Geoffrey E. Hinton for their foundational work in machine learning with artificial neural networks, underscores the fundamental role of physics in driving technological innovations.
We argue that a significant number of AI developers in the medical imaging field lack a background in physics, necessary for the correct understanding of medical images, which is essential to build trustworthy AI systems in healthcare. Physics-based methods integrated in AI algorithms provide enhanced reliability and system safety by accurately representing the underlying physical relationships in medical images.
In this work, we aim to offer a pedagogical handbook on the basic physical principles of medical imaging for every existing modality in the clinics, that can serve for consultation, and provide references for the reader interested in further study. In addition, we connect each modality with the latest trends in AI for image generation, reconstruction, and physics-informed algorithms. 
We build upon Bushberg \textit{et al.} \cite{bushberg2021essential}, which was our reference for the physical principles of medical images, and we synthesize the basis of physics necessary to build trustworthy AI algorithms in medical imaging. We highlight the challenges and relate the reader to the current state-of-the-art in each modality, to finally conclude with an overview of physics-informed machine learning to enhance AI systems in medical imaging. The main contributions of this paper can be summarized as follows:
\begin{itemize}
    \item A review of the fundamental physics behind every clinical imaging modality, serving as a structured and accessible resource for AI researchers.
    \item A compilation of recent advances in AI and physics-informed algorithms for medical imaging applications. The contributions were selected from relevant journal and conference papers from prestigious publishers such as “Elsevier”, “IEEE”, “Springer”, “Nature”, etc.
    \item An overview of physics-informed machine learning, discussing how integrating physics constraints into machine learning models can improve interpretability, robustness, and clinical applicability.
    \item A discussion of challenges and limitations of AI in medical imaging in relation with the potential of leveraging physics in AI algorithms, current and future trends in physics-informed machine learning, and their impact on developing trustworthy medical imaging.
\end{itemize}
The remainder of this survey is structured as follows: Section \ref{sec:physics_medical_images} introduces the physics behind each medical imaging modality existing in the clinics (Section \ref{sec:vis_im}-\ref{sec:combined}), as well as image quality challenges (Section \ref{sec:challenges}). Section \ref{sec:physics_informed}, presents physics-informed machine learning algorithms, detailing existing approaches (Section \ref{sec:obs_biases}-\ref{sec:ind_biases}), and challenges (Section \ref{sec:physics_informed_challenges}). Finally, Section \ref{sec:concl} discusses the potential of physics-informed models, future trends and conclusions.

\section{Physics behind medical imaging by modality}
\label{sec:physics_medical_images}
\begin{figure}
    \centering
    \includegraphics[page=1,angle=0,trim={8.4cm 1.3cm 7.5cm 2.3cm},clip, width=\linewidth]{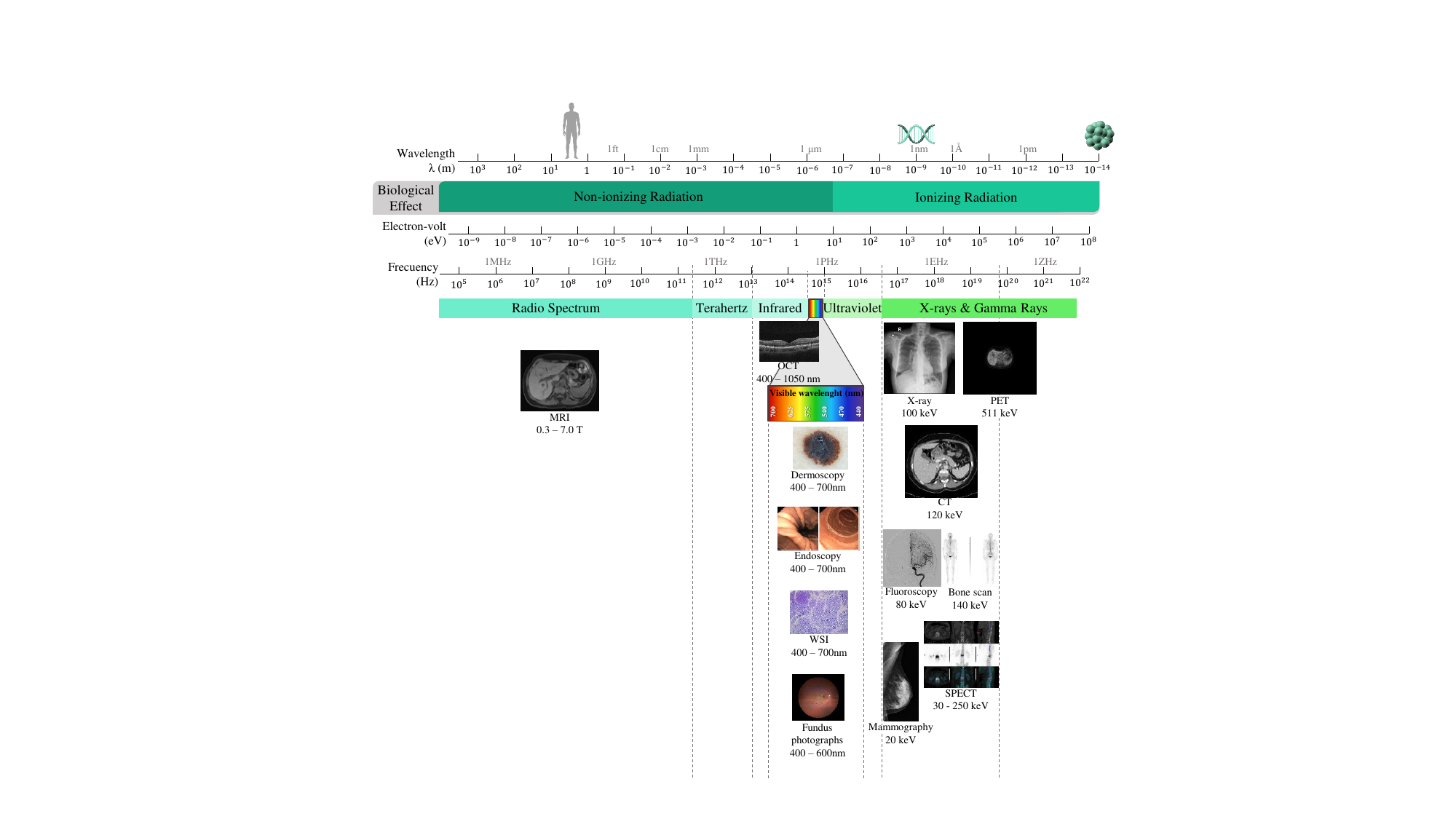}
    \caption{Medical imaging modalities in the electromagnetic radiation spectrum.}
    \label{fig:spectrum}
\end{figure}

\begin{figure}
    \centering
    \includegraphics[page=2,angle=0,trim={16.8cm 10.8cm 7.5cm 2.3cm},clip, width=0.64\linewidth]{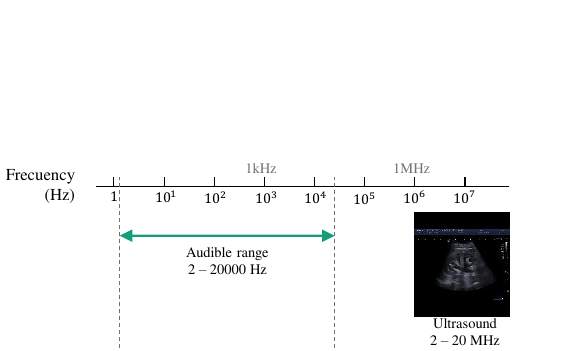}
    \caption{Ultrasound imaging in the acoustic spectrum.}
    \label{fig:ultrasound}
\end{figure}

Radiation is the propagation of energy through space or matter. Electromagnetic radiation, subatomic particle radiation (nuclear imaging), and acoustic radiation (ultrasound imaging) are the main types of radiation for medical imaging. Figure \ref{fig:spectrum} depicts the electromagnetic radiation spectrum for the different medical imaging modalities available (note that radiation ranges in ionizing imaging techniques are approximate and depend on the patient's size). Figure \ref{fig:ultrasound} shows ultrasound imaging in the acoustic spectrum. In the following paragraphs, we introduce all the existing clinical medical imaging modalities shown in Figure \ref{fig:spectrum} and \ref{fig:ultrasound}.

Given the variety of imaging techniques and the importance of consistency in medical diagnostics, it becomes crucial to have a standard for managing and sharing medical images. This is where DICOM (Digital Imaging and Communications in Medicine) comes into play. DICOM is the universal standard that defines and controls the formats for sending, distributing, and storing medical images across different machines, manufacturers, and imaging modalities \cite{pianykh2012digital}. It plays a central role in ensuring that images from various sources can be accessed, interpreted, and analyzed consistently, regardless of the technology used. DICOM is implemented in nearly all radiology, cardiology, and radiotherapy devices, and its use is expanding to other medical fields such as ophthalmology and dentistry \cite{dicom_webpage}.

\subsection{Visible spectrum images}
\label{sec:vis_im}

Visible light is utilized to produce 2D images or videos in fields such as dermatology, gastroenterology, histology and ophthalmology. In dermatology, the most widely used technique to capture skin images is dermoscopy (a dermatoscope is simply a magnifying lens), followed by  total-body digital photography \cite{schneider2019emerging}. Endoscopy uses visible light to illuminate different parts of the gastrointestinal tract, capturing images or videos of the structures of interest. Examples of applications of endoscopy are colonoscopy or laparoscopy, yet its utility spans a broad spectrum of medical procedures. Current trends in AI applications in endoscopy can be found in \cite{chahal2020primer}. Histology relies on visible light microscopy to examine tissue and cells stained with specific dyes at different magnifications. The introduction of whole slide imaging (WSI) in 1999 revolutionized pathology by enabling high-resolution digital slides \cite{kumar2020whole}. Bahadir \textit{et al.} review the latest trends of AI in histopathology \cite{bahadir2024artificial}. Although light microscopy is the main diagnostic tool in histology, Transmission Electron Microscope is performed routinely on renal biopsies. In this context, recent work by Zhang \textit{et al.} \cite{zhang2023deep} applied DL techniques on electron microscopy images of renal biopsy. In ophthalmology, color fundus photography employs a fundus camera to record color images of the retina (fundus). Diverse applications of AI are reviewed in \cite{grzybowski2024retina}, and the latest studies showed that fundus photographs can be used to monitor the progression of neurodegenerative disorders. 

\subsubsection{Optical Coherence Tomography}

Optical Coherence Tomography (OCT) was developed in the 1990s for non-invasive cross-sectional imaging in biological systems \cite{huang1991optical}. Since then, it is evolving from near-infrared illumination to visible light optical coherence tomography \cite{wang2024dual}, demonstrating its effectiveness in preclinical and ophthalmic imaging. OCT holds great promise for AI applications \cite{Kalupahana_OCT_IEE_Access}, with some recent advances in the field of neurological diseases in relation to OCT \cite{alvarez2024fully}. 

\subsection{X-ray imaging}

X-rays are most likely the best known medical imaging modality. The underlying physical principle is simple, and yet effective: X-rays are a type of electromagnetic radiation produced by high energy electrons. These electrons are produced due to the ionization of nitrogen and oxygen atoms, which attract positive ions to the cathode, and therefore inject electrons that are accelerated to the anode. The resulting X-rays can be directed to a patient, and then collected in a detector. X-rays are differently absorbed by tissues, which allows to capture their interactions with the patient's anatomy in a radiographic image. There are four major types of interactions of photons with matter, but only three of them play a role in diagnostic radiology and nuclear medicine (Rayleigh scattering, Compton scattering and photoelectric effect) \cite{bushberg2021essential}, while the last one (pair production, can only occur when the energy of photons exceeds 1.02 MeV) has only been simulated at a theoretical level for monitoring of radiotherapy dosing \cite{lyu2023tomographic}.

\subsubsection{Radiography}

Radiography was the first medical imaging technology. This technique captures the attenuation (absorption and scattering) of an homogeneous distribution of X-rays entering a patient, which is then modified by the interactions with the different tissues, resulting in an heterogeneous distribution emerging from the patient that is recorded in a 2D radiograph \cite{bushberg2021essential}. The kilovoltage, X-ray exposure time, and beam size are adjusted according to the anatomical area under study. In the field of dentistry, individual dental radiographs and orthopantomograms are performed. Orthopantomograms are panoramic dental radiographs produced by rotating the X-ray tube around the patient’s head, generating a comprehensive 2D image of the dental and maxillofacial structures. \cite{rozylo2021panoramic}. Advances in artificial intelligence (AI) have further enhanced the interpretation of these images, as reviewed by Costa \textit{et al.} \cite{alharbi2024exploring}.

\subsubsection{Computed Tomography}

Computed Tomography (CT) images are 3D images generated by producing multiple (around 1000) X-ray projection images across a broad angular range, typically 360\degree, rotating the X-ray tube and detector around the patient \cite{bushberg2021essential}. Simultaneous rotation of the X-ray source and translatory movement of the patient allow to achieve continuous data acquisition throughout the volume of interest \cite{brink1994helical}. This geometry corresponds to an helical CT scanner, which represents the vast majority of CT scanners in use today.  Previous to helical CT scanners, there were sequential CT scanners, which are still in use for cranial imaging \cite{van2007image}. In sequential CT scanners the patient remains stationary during each full rotation of the X-ray tube, advancing incrementally to acquire axial slices. This technique enhances image resolution but results in increased radiation exposure and scanning time compared to helical CT. The generation of the 3D image relies on advanced reconstruction algorithms, with the resulting voxel values represented in grayscale, ranging from -1000 to 1000. The grayscale in CT is named Hounsfield Unit (HU), after one of the main developers of this technology \cite{bushberg2021essential}. X-ray CT scanners typically operate at 120 kV, however, the voltage can be optimized based on the specific application and the patient's size \cite{bushberg2021essential}. Deep Learning (DL) has shown potential to personalize optimization of imaging protocols to minimize radiation exposure while maintaining clinical image resolution \cite{salimi2022deep}. 

Spatial resolution depends on several factors including physics related aspects such as X-ray focal spot size, number of projection views per rotation of the X-ray tube, detector cell size; together with reconstruction algorithms \cite{wang2018improving}.
Until 2009, the lack of computational power prevented the clinical introduction of iterative algorithms for image reconstruction in CT, which rapidly replaced filtered back projection (FBP). In the latter, CTs were reconstructed from projections (sinograms) by applying a high-pass filter followed by a backward projection step \cite{willemink2019evolution}. The main drawback of FBP is the significant reduction in image quality when radiation dose is decreased, due to the increase in image noise. Recently, AI has emerged as a new promising technique to improve CT image reconstruction, showing potential to reduce CT radiation doses while speeding up reconstruction times \cite{willemink2019evolution, singh2020artificial, brady2023implementation, bushberg2021essential}. However, DL-based reconstruction algorithms require large training datasets, and are prompt to biases in different subpopulations if training data significantly differs from target population.

There are particularities of CT depending on the clinical application.
For lung cancer screening \cite{national2011reduced}, Low-Dose Computed Tomography (LDCT) utilizes a lower dose of radiation to scan the patient, which is achieved lowering the X-ray flux.  However, lowering the dose of radiation in LDCT increases image noise and, thus, reduces signal-to-noise ratio (SNR) and image quality. Physics-/model-based data-driven methods for LDCT are surveyed by Xia \textit{et al.} \cite{xia2023physics}. 

The incorporation of contrast agents, which facilitate the evaluation of patient hemodynamics and the characteristic vascularization of tissues, has enabled the development of various advanced acquisition modalities, including perfusion imaging, which assesses blood flow dynamics, aiding in the diagnosis of stroke and myocardial perfusion, virtual CT colonoscopy, which generates 3D images of the colon providing a non-invasive alternative for detecting polyps and tumors, or prospectively gated cardiac CT, which reduces motion artifacts by synchronizing image acquisition with the cardiac cycle, enhancing coronary artery evaluation \cite{bushberg2021essential, greffier2024photon}.

Next-generation modern CT systems have integrated dual-energy technology (DECT), wherein X-ray spectra are captured at both low and high energy levels. This approach enables the independent assessment of the contributions from photoelectric effect and Compton scattering. DECT facilitates material decomposition by leveraging differences in attenuation coefficients at varying energy levels, thereby distinguishing materials with similar HU but differing atomic compositions. This technology can produce various image types, such as virtual monoenergetic images, material-specific images (e.g., iodine maps), and virtual non-contrast images, significantly enhancing tissue characterization and lesion detection capabilities \cite{odedra2022dual}. Recent advancements in generative AI have further improved these capabilities. Jeong \textit{et al.} \cite{jeong2023synthetic} explored the application of generative AI techniques to enhance DECT imaging, focusing on improving image quality and diagnostic accuracy through machine learning methods. 

Additionally, as an emerging technology with the potential to change clinical CT, photon-counting CT (PCCT) integrates new energy-resolving X-ray detectors to count the number of incoming photons and measure their energy, resulting in higher contrast-to-noise ratio, improved spatial resolution, and optimized spectral imaging at a lower radiation exposure \cite{willemink2018photon}. In conventional CT, finer detector space leads to larger datasets that allow to achieve higher resolution, increasing reconstruction time, which is not required for many clinical tasks \cite{bushberg2021essential}. PCCT energy-resolving detectors eliminate electronic noise, in comparison with the traditional energy-integrating detectors in CT. Greffier \textit{et al.} \cite{greffier2024photon} comprehensively review PCCT and compare their technical innovations against conventional CT. 
Nevertheless, as any innovative technology, PCCT also encounters challenges, such as detector charge-sharing effects or Compton scattering, which may lead to errors during the reconstruction process that degrade image quality. Yu \textit{et al.} \cite{yu2024material} proposed a novel physics-guided material decomposition model for PCCT, that leverages DL and incorporates critical physics parameters, which underlines the role of physics in building reliable DL systems in medical imaging.

\subsubsection{Mammography}

Mammography is an optimized radiography examination specifically designed for detecting breast cancer at an early stage. Screening mammography attempts to detect breast cancer in the asymptomatic population, while diagnostic mammography aims to assess and delineate lesions identified by the former \cite{bushberg2021essential}. AI has holds significant potential for improving breast cancer screening, current state of the art and challenges are reviewed by Díaz \textit{et al.} \cite{diaz2024artificial}.

Modern mammography systems incorporate rotating X-ray tubes that facilitate tomosynthesis, a technique grounded in the same physical principles as sequential CT. This advancement enables the acquisition of more detailed and comprehensive imaging, offering enhanced diagnostic information in appropriately selected cases. Additionally, the introduction of contrast-enhanced mammography, combined with DL algorithms, has shown diagnostic performance comparable to magnetic resonance imaging (MRI), especially for evaluating dense breast tissue \cite{sorin2024deep}. Furthermore, in breast imaging, ultrasound (US) is used as a supplemental screening alternative for women with dense breast tissue \cite{bushberg2021essential}, and recently MRI screening has been recommended for women with extremely dense breasts \cite{mann2022breast}.

\subsubsection{Fluoroscopy}

Fluoroscopy shows real-time X-ray imaging of internal anatomic structures for the placement of medical devices, such as catheters and stents, or the observation of temporal physiological phenomena in patients  providing a dynamic display of the structure of interest \cite{bushberg2021essential} . Fluoroscopy systems can operate in two modes: (1) fluoroscopy, real-time imaging for positioning, which is usually not recorded and involves relatively low radiation exposure, and (2) fluorography, which records clinically relevant sequences using a pulsed radiographic mode, giving higher radiation levels, similar to radiographic imaging \cite{bushberg2021essential}. The use of contrast media is critical in fluoroscopy for enhancing the visibility of internal structures and improving diagnostic accuracy. Contrast agents, such as iodine-based compounds or barium sulfate, allow for the differentiation of soft tissues, blood vessels, and other anatomical features that would otherwise be indistinguishable in conventional X-ray imaging. In procedures such as angiography, the injection of iodinated contrast highlights the vascular system, enabling clinicians to visualize blood flow, detect blockages, and guide interventions with precision. Similarly, in gastrointestinal studies, barium-based contrast delineates the digestive tract, facilitating the assessment of structural and functional abnormalities \cite{lopez2024fluoroscopy} AI applications are more complex for interventional radiology (IR) than for diagnostic radiology. IR encompasses preprocedural diagnostic imaging, procedural imaging guidance, prosprocedural imaging evaluation, and therapeutic tools, mostly relying on unstructured data, which challenges AI applications. The potential of AI in the field of IR is reviewed by Glielmo \textit{et al.} \cite{glielmo2024artificial}.

\subsection{Magnetic Resonance Imaging}


Magnetic resonance imaging (MRI) studies the magnetic properties of the nucleus of the atom through radio-frequency (RF) waves. The atomic nucleus is constituted of protons and neutrons, which exhibit a magnetic field associated with their nuclear spin and charge distribution. In contrast with X-ray imaging, MRI does not employ ionizing radiation, however, it is a more expensive imaging modality compared to the former. The key components are the magnet, magnetic field gradient coil, and RF coils. In MRI, a strong external magnetic field generated by the magnets causes the individual nuclei to selectively absorb, and then release, energy unique to those nuclei and their surrounding environment. This energy coupling is known as resonance \cite{bushberg2021essential}. An RF coil is basically a resonant circuit, which is tuned to the resonance frequency of proton spins for a given magnet field (similar to a radio tuned to the frequency of a radio station) \cite{kwok2022basic}. The typical magnetic field strengths for MRI systems range from $0.3$ to $7.0$ T, which require the electromagnet core wires to be superconductive \cite{bushberg2021essential}. Magnetic field gradients are essential to localize signals generated during MRI process, as these fields interact with the main (and much stronger) magnetic field \cite{bushberg2021essential}. 

MRI sequences leverage the resonance to generate images with varying contrasts based on tissue properties, in which each voxel (3D pixel representing a small unit of volume in an image) depends on the number of protons (proton density) and magnetic properties of the tissue in that voxel. MRI can produce high contrast images due to the different local magnetic field properties of different types of tissue (fat, white and gray matter in the brain, cancer, etc.) \cite{bushberg2021essential}. The two primary sequences, T1-weighted and T2-weighted imaging, provide distinct diagnostic information. T1-weighted imaging focuses on the longitudinal relaxation of protons, using short repetition time (TR) and short echo time (TE), where tissues with short T1 times appear bright and fluids appear dark, offering excellent anatomical detail. In contrast, T2-weighted imaging emphasizes transverse relaxation, employing long TR and long TE, where tissues with long T2 times and fluids appear bright, making it particularly effective for identifying pathology, inflammation, and edema. Additional sequences include Proton Density (PD), which highlights proton concentration differences using long TR and short TE; FLAIR (Fluid-Attenuated Inversion Recovery), which suppresses cerebrospinal fluid (CSF) signals to enhance lesion detection near CSF spaces; and Diffusion-Weighted Imaging (DWI), which assesses water molecule diffusion, and is critical for early stroke diagnosis. \cite{brown2014magnetic}. MRI can also monitor blood flow in arteries (MR angiography), and blood flow in brain (functional MR) \cite{bushberg2021essential}. 

MRI data are initially stored in the raw spatial frequency domain, the \textit{k}-space matrix. The \textit{k}-space encodes spatial frequency values in a four quadrant 2D matrix of complex values, where the origin at the center represents frequency zero, the central region contains lower spatial frequencies, and the higher spacial frequencies are in the periphery  \cite{bushberg2021essential}. Each point in the \textit{k}-space corresponds to a specific spatial frequency component of the final image. The \textit{k}-space matrix is filled one row at a time in a conventional acquisition, by systematically collecting data points through the application of gradient magnetic fields, which encode spatial information by varying the frequency and phase of the detected signals \cite{bushberg2021essential}. Once rows in the \textit{k}-space matrix are fully populated, image reconstruction is performed using the inverse fast Fourier transform (FFT). This mathematical operation decodes the frequency-domain data in the \textit{k}-space matrix to produce the spatial domain representation, revealing the anatomical structures of the scanned area. The final image is processed to represent photon density, T1, T2 and flow characteristics of the tissues using a grayscale range, with each pixel corresponding to a voxel \cite{bushberg2021essential}. The organization and density of data sampling in the \textit{k}-space matrix directly influence image quality, resolution, and the presence of artifacts. 

Image reconstruction in MRI is influenced by the physical effects that are included in the signal model \cite{fessler2010model}, and presents challenges due to long acquisition times. Several research efforts have focused on accelerating MRI, i.e., developing methods to reconstruct images from under-sampled data \cite{heckel2024deep}. Traditional reconstruction methods from under-sampled \textit{k}-space data include parallel imaging and compressed sensing, widely used in the clinics \cite{kim2024review}, although both encounter practical limitations \cite{heckel2024deep}. DL methods have enabled transforming under-sampled or noisy data into high quality images, mitigating artifacts and accelerating the imaging process \cite{heckel2024deep}. Generative adversarial neural networks (GANs) have been explored in MRI reconstruction to estimate missing \textit{k}-space samples, and fix artifacts in the image-space \cite{shaul2020subsampled}. In a recent proof of concept, Okoli \textit{et al.} \cite{okolie2024accelerating} proposed a score-based diffusion model for accelerating MRI reconstruction, although they underscored the need of further research and clinical assessment. Furthermore, model-based methods considering physical effects and integrating neural networks have shown potential to improve image quality \cite{fessler2010model, xin2023fill}. Another recent study by Peng \textit{et al.} \cite{peng2022learning} proposed a \textit{k}-space acquisition optimization strategy conditioned on MRI physics for accelerated MRI reconstruction using a Neural Ordinary Differential Equation (ODE) combined with DL-based reconstruction. These works pave the way for further research in physics-based algorithms.

Patient motion represents a significant challenge in MRI, where acquisition times range from 20 to 60 minutes depending on the different sequences added. Quantitative MRI (qMRI) derives physical tissue properties from a set of qualitative images
captured with different imaging settings, facilitating consistent measurement of biomarkers, in spite of longer acquisition times. In contrast to conventional MRI, which relies on relative signal intensities for visual interpretation, qMRI quantifies parameters such as T1 and T2 relaxation times, proton density, and diffusion coefficients in standardized units. This facilitates reproducible comparisons across subjects, scanners, and time points, while minimizing hardware-related variability. By employing multiple acquisitions with differing parameters, qMRI enhances measurement precision and specificity, allowing for the detection of subtle changes in tissue integrity and composition, such as myelin content and iron concentration \cite{shah2011quantitative,   weiskopf2021quantitative}. Recent work by Eichhorn \textit{et al.} \cite{eichhorn2024physics} proposed physics-informed motion correction (through a physics-informed loss) to leverage information from the MRI signal evolution to detect and exclude motion-corrupted \textit{k}-space lines from a data consistent reconstruction. 

MRI also employs contrast agents to enhance the visualization of internal structures during imaging examinations. These agents are typically administered intravenously, with gadolinium-based contrast agents (GBCAs) being the most commonly utilized. GBCAs primarily shorten the T1 relaxation time of tissues, leading to increased signal intensity on T1-weighted images. Additionally, iron oxide-based agents, including super-paramagnetic iron oxide (SPIO) and ultra-small super-paramagnetic iron oxide (USPIO), are utilized to reduce T2 signal intensity, while manganese-based agents enhance T1 signal intensity. The selection of a specific contrast agent depends on the clinical application, with some agents designed for targeted organ imaging, such as liver-specific contrast agents \cite{xiao2016mri}.

\subsection{Nuclear images}

Nuclear medicine is a specialized field of medical imaging that uses radioactive isotopes, known as radiotracers, to visualize physiological and metabolic processes within the body. These radiotracers may be bound or unbound to other molecules and interact at the cellular level, emitting ionizing radiation during nuclear decay. The emitted radiation consists of subatomic particles, such as alpha particles (helium nuclei) or beta particles (electrons or positrons), or gamma rays (photons). This radiation is detected to generate images that reflect metabolic activity \cite{bushberg2021essential}. While these images offer valuable functional information, they typically present lower anatomical resolution compared to other imaging modalities. Advances in nuclear medicine have led to the development of two primary imaging techniques: Single-Photon Emission Computed Tomography (SPECT) and Positron Emission Tomography (PET), each offering unique diagnostic capabilities. Both SPECT and PET rely on the principle that ionizing radiation emitted by radiotracers traverses anatomical structures with varying attenuation based on tissue density \cite{crișan2022radiopharmaceuticals}. These techniques provide critical insights into physiological and metabolic processes, playing a pivotal role in the diagnosis and management of numerous conditions, including cardiovascular diseases, neurological disorders, and cancers \cite{enlow2023state}.
However, nuclear medicine techniques have several limitations, including lower spatial resolution compared to CT or MRI, which can hinder the detection of small structures \cite{cherry2009multimodality}. The use of ionizing radiation requires careful dose management to minimize patient risk. Additionally, the short half-lives of radiotracers pose logistical challenges related to their production, distribution, and timely administration, while the high cost and limited availability of radiotracers can restrict access to imaging services \cite{enlow2023state}. Despite these challenges, nuclear medicine remains a crucial tool in modern diagnostics due to its ability to provide unique functional and metabolic information.

\subsubsection{Single Photon Emission Computed Tomography}
Single-Photon Emission Computed Tomography (SPECT) is a nuclear medicine imaging technique that provides 3D functional information about biological processes within the body. SPECT imaging relies on gamma-emitting radiotracers, such as technetium-99m (Tc-99m), which is the most commonly used isotope due to its favorable energy characteristics and high half-life  \cite{rathmann2019radiopharmaceutical}. 
Tc-99m decays to a more stable form, Tc-99, by emitting gamma rays (140 keV \cite{peterson2011spect}) which pass through the body interacting with tissues according to their density and the radiation's penetration power. High-density structures, such as bones, attenuate more radiation than lower-density tissues, such as fat, a difference that is captured by detectors to produce images. Gamma cameras detect these photons using collimators that allow only photons traveling at specific angles to pass through, ensuring precise image formation \cite{bushberg2021essential}. By acquiring multiple 2D projections from different angles, tomographic reconstruction algorithms create detailed 3D images that enhance lesion localization and characterization. SPECT’s versatility is demonstrated by its wide range of clinical applications as myocardial perfusion imaging, cerebral blood flow, pulmonary ventilation/perfusion (V/Q), tumor detection or bone scintigraphy. Notably, SPECT bone scintigraphy plays a critical role in evaluating bone metabolism, facilitating the identification of fractures, infections, and metastases \cite{mannarino2023radionuclide, ballinger2021radiopharmaceuticals}.

\subsubsection{Positron Emission Tomography}
Positron Emission Tomography (PET) is an advanced molecular imaging technique that visualizes biological functions and metabolic processes with high sensitivity. PET relies on radiotracers labeled with positron-emitting radionuclides. The most commonly used radiotracer is fluorine-18-labeled fluorodeoxyglucose (18-FDG), which mimics glucose metabolism. As Fluorine-18 decays to Oxygen-18, it emits a positron that annihilates upon encountering an electron, producing two 511 keV gamma photons traveling in opposite directions. PET scanners detect these photon pairs using rings of detectors in a process known as coincidence detection, which allows precise localization of the decay event \cite{bushberg2021essential}. Advanced image reconstruction algorithms then generate 3D images of the tracer distribution within the body, mapping metabolic activity at a molecular level. 

PET imaging is highly versatile, employing various radiotracers for different clinical applications. 18-FDG is primarily used in oncology for cancer detection, staging, monitoring treatment response, and assessing recurrence \cite{casali2021state}. In neurology, PET evaluates brain function in conditions such as dementia, epilepsy, and neurodegenerative diseases \cite{tan2020total}. In cardiology, PET assesses myocardial perfusion and viability, aiding in the planning of coronary artery bypass graft procedures. Additionally, radiotracers such as Carbon-11, Nitrogen-13, and Oxygen-15 are utilized to study specific metabolic processes, while specialized tracers are designed for particular cancers or neurological disorders \cite{tan2020total}. Hagos \textit{et al.} \cite{hagos2024recent} underscored the potential of generative AI and large language models to advance nuclear medicine practices, offering new avenues to improve diagnostic accuracy and workflow efficiency.

\subsection{Ultrasound}
Ultrasound (US) waves are mechanical high frequency sound waves that require an elastic medium to spread over, unlike the previously described medical imaging modalities that leverage electromagnetic (EM) waves. Sound waves are longitudinal waves, oscillating in the direction of travel. In contrast, EM waves are transverse waves, oscillating perpendicular to the direction of travel \cite{patey2021physics}. 
In the process, an ultrasound transducer, that is in direct physical contact with the patient, generates a short-duration pulse of sound, which travels into the tissue and is reflected by internal structures and organs in the body, such that the transducer receives and records the amplitude and time delay of the echoes for a given direction of the pulse \cite{bushberg2021essential}. The repetition of the process across a wider area in the anatomical area of interest allows to create an US image. In modern US transducers, multiple elements can transmit and receive the pulses resulting in a brightness mode (B-mode) ultrasound image of a planar section of tissues \cite{bushberg2021essential}. 
Apart from the B-mode, there are A-mode and M-mode. A-mode (amplitude) is the processed echo amplitude versus time generated as output by the receiver, the simplest mode of US generation by a single transducer \cite{bushberg2021essential, patey2021physics}. It was the earliest application of US in medicine, and now it is almost obsolete. However, A-mode is sometimes combined with M-mode imaging, it is also used in ophthalmology applications for accurate measurements of the eye, and it is utilized in therapeutic US \cite{bushberg2021essential, patey2021physics}. M-mode (motion), also known as T-M mode (time-motion), is commonly used in echo-cardiography. It employs B-mode information from a stationary US beam to track velocities of echoes generated from moving structures acting as reflectors throughout the cardiac cycle \cite{bushberg2021essential, patey2021physics}.

Additionally, Doppler modes are essential for evaluating blood flow. Color Doppler displays blood flow direction and velocity, while Pulse Wave and Continuous Wave Doppler measure the speed of blood flow or tissue movement, leveraging the Doppler effect to detect changes in frequency due to motion \cite{bushberg2021essential}.
Optimizing US image quality involves selecting appropriate settings for the specific anatomical area being examined. To facilitate this process, manufacturers provide preset configurations tailored to different target areas, simplifying image acquisition for operators who may not have extensive knowledge of the underlying physics. Once the appropriate preset is selected, several key parameters can be adjusted to fine-tune image quality. Brightness is influenced by gain settings, including Overall Gain and Time Gain Compensation (TGC), which compensates for signal attenuation at different depths. Lateral Gain Compensation (LGC) adjusts brightness horizontally. Proper gain settings ensure that low echogenicity structures, such as fluids, appear black, while highly echogenic structures, like bones, appear white. Other critical parameters include Depth, Dynamic Range, Focal Zone, and Frequency. Higher frequencies provide better resolution but lower penetration, while dynamic range settings influence contrast and visibility of details \cite{zander2020ultrasound}. 
To enhance image quality, Tissue Harmonic Imaging (THI) is employed. This technique relies on the non-linear propagation of US waves through tissue, resulting in the generation of harmonic frequencies (multiples of the fundamental frequency). The second harmonic frequency is typically used for image formation, as higher harmonics are attenuated. THI improves image resolution, reduces artifacts, and improves the visualization of deeper structures \cite{anvari2015primer}.

Building on the principles of harmonic imaging, Ultrasound Contrast Agents (UCAs), administered intravenously, further enhance diagnostic capabilities by improving the visualization of blood flow and tissue perfusion. UCAs are gas-filled microbubbles, typically 1-8 $\mu m$ in diameter, stabilized by a phospholipid or protein shell. When exposed to US waves, these microbubbles oscillate and resonate, producing strong echoes due to their nonlinear behavior. This resonance generates harmonic frequencies that can be selectively detected, significantly increasing the contrast-to-tissue ratio. Techniques like harmonic imaging and pulse inversion exploit these properties to differentiate microbubble signals from background tissue. UCAs are employed in cardiology to define endocardial borders and assess myocardial perfusion, as well as in radiology to characterize liver lesions \cite{yusefi2022ultrasound}.

Optimizing US image quality involves adjusting several key parameters. Brightness is influenced by gain settings, including Overall Gain and Time Gain Compensation (TGC), which compensates for signal attenuation at different depths. Lateral Gain Compensation (LGC) adjusts brightness horizontally. Proper gain settings ensure that low echogenic structures, such as fluids, appear black, while highly echogenic structures, like bones, appear white. Other critical parameters include Depth, Dynamic Range, Focal Zone, and Frequency. Higher frequencies provide better resolution but lower penetration, while dynamic range settings influence contrast and visibility of details \cite{image2007applied, zander2020ultrasound}.

In quantitative ultrasound imaging, the goal is to quantify interactions between US and biological tissues, enhancing diagnostic capabilities \cite{cloutier2021quantitative}. Ultrasound elastography extends these applications by evaluating tissue stiffness, aiding in differential diagnoses and identification of biopsy targets. This combination of modes, techniques, and image optimization parameters makes US a versatile and indispensable tool in modern medical imaging, complementing other diagnostic modalities \cite{hasegawa2021advances}. 

Recent work in US explored improving the quality of AI generated US images by introducing a physics-based diffusion model specifically designed for this modality \cite{dominguez2024diffusion}. The proposed customized noise scheduler simulated the attenuation of echoes returning to a US receiver. This work opens the door to further refinements based on the physics of US, such as reflection and scattering.

\subsection{Combined imaging modalities}
\label{sec:combined}

Combined imaging modalities integrate multiple imaging techniques to provide comprehensive anatomical and functional information, enhancing diagnostic accuracy and clinical decision-making. Leveraging advanced computational techniques and DL-based algorithms, combined imaging modalities improve image quality, diagnostic specificity, and workflow efficiency. These techniques address the limitations of single-modality imaging, providing clinicians a more holistic view of patient anatomy and physiology, ultimately improving patient outcomes across various medical specialties \cite{huang2020review,hussain2024recent,furtado2023synergistic}. These methods encompass both hardware-based hybrid systems and software-based fusion techniques, each offering distinct advantages and applications in medical diagnostics.

\subsubsection{Hardware-based combined imaging modalities}

Hardware-based hybrid systems combine two imaging modalities into a single device, enabling simultaneous or near-simultaneous image acquisition. Notable examples include PET/CT, SPECT/CT, and PET/MRI. PET/CT merges the metabolic insights of PET with the high-resolution anatomical detail of CT, demonstrating its critical role in cancer detection, staging, and treatment monitoring. SPECT/CT combines functional SPECT data with CT imaging for improved attenuation correction and localization, widely applied in cardiology, oncology, and neurology \cite{cherry2009multimodality}. PET/MRI integrates PET’s molecular imaging capabilities with MRI's superior soft tissue contrast, providing reduced radiation exposure and enhanced diagnostic capabilities for neuroimaging and oncology \cite{beyer2009decade}. These hybrid systems minimize patient movement artifacts, improve image co-registration, and streamline clinical workflows, resulting in more accurate and efficient diagnoses.

Recent work by Sudarshan \textit{et al.} \cite{sudarshan2021towards} proposed a deep neural network uncertainty-framework to predict standard-dose PET images from a combination of low-dose PET images and multi-contrast MRI (acquired during simultaneous PET-MRI), proposing a transform-domain loss inspired in the physics image acquisition process, that modeled the underlying sinogram-based physics of the PET imaging system. Their approach enhanced the robustness to unseen out-of-distribution (OOD) acquisitions that differed from the training set distribution, which could arise from variations in radiotracers, anatomy, pathology, photon counts, hardware, and reconstruction protocol \cite{sudarshan2021towards}.

\subsubsection{Software-based combined imaging modalities}

Software-based fusion techniques involve the computational combination of images obtained separately from different modalities. Examples include dual-energy CT (DECT) fusion, where images captured at different energy levels produce virtual monoenergetic images, material decomposition maps, and virtual non-contrast images \cite{huang2020review}. AI-based algorithms, such as convolutional neural networks (CNNs), further enhance DECT fusion by improving material differentiation and image quality \cite{liang2024medical}. 
Zhao \textit{et al.} \cite{zhao2020dual} developed a deep learning approach for DECT imaging that predicts high-energy images from low-energy data, facilitating accurate virtual non-contrast imaging and iodine quantification. Similarly, Li \textit{et al.} \cite{li2023improved} proposed an iterative neural network incorporating CNNs for high-quality image-domain material decomposition, demonstrating superior performance over traditional methods. These advancements underscore the potential of AI to enhance DECT imaging, offering more detailed and reliable diagnostic information. MRI fusion techniques combine sequences like T1-T2 fusion for enhanced tissue contrast and diffusion-perfusion fusion for stroke evaluation \cite{liang2024medical}. Multi-modality fusion methods, such as CT-MRI, PET-CT, and SPECT-CT, integrate functional and anatomical data, offering precise diagnosis and improved treatment planning. Additionally, ultrasound-MRI fusion supports real-time image guidance for procedures like biopsies \cite{khorasani2023performance}.

\subsection{Image quality challenges: artifacts and technical limitations}
\label{sec:challenges}
Medical image quality is critical for the precise and reliable development of DL algorithms in diagnostic and analytical processes. Several inherent challenges in medical imaging modalities, such as artifacts, technical limitations, and data heterogeneity, have a substantial impact on diagnostic outcomes, as will be explained in the next paragraphs.

\subsubsection{Artifacts}

Artifacts in medical imaging are unintended distortions or errors which can compromise image quality and hinder effective interpretation. Such artifacts may arise from a variety of factors, including patient movement, limitations in imaging physics, hardware or software issues, and image processing techniques \cite{bushberg2021essential}. Patient movement, such as involuntary motion during scans, is a common source of motion artifacts, leading to blurring or ghosting effects that degrade the quality of MRI and CT images \cite{barrett2004artifacts}. Imaging physics constraints, such as the differential absorption of X-ray photons by tissues of varying density, result in beam-hardening artifacts in CT, which manifest as streaking patterns that obscure diagnostic details \cite{barrett2004artifacts, bushberg2021essential}. Hardware limitations, such as imperfections in imaging systems or detectors, and software errors during data reconstruction can also introduce distortions \cite{bushberg2021essential, barrett2004artifacts}. Additionally, image processing techniques, including under-sampling or compression, can lead to aliasing artifacts in MRI, where high-frequency signals are misrepresented as lower frequencies, complicating accurate image interpretation \cite{bushberg2021essential}. These diverse sources highlight the multifactorial nature of artifacts and their significant impact on diagnostic accuracy.

The latest advances in DL and generative AI have shown potential in reducing artifacts and enhancing image quality. CNNs, GANs and diffusion models have been successfully employed for artifact removal, de-noising and synthetic data generation \cite{xu2022research}. Recent novel approaches, including StylEx \cite{lang2024using} and Dual-Domain Optimization \cite{amirian2024artifact}, further address challenges such as model interpretability and edge artifacts, improving the overall reliability of diagnostic imaging.

\subsubsection{Technical limitations}

Technical limitations of imaging modalities and physical protocols are another source of degradation in image quality. Resolution constraints, such as limited spatial resolution, hinder the detection of small lesions or fine anatomical structures, particularly in modalities like ultrasound or LDCT imaging \cite{bushberg2021essential}.
Similarly, insufficient voxel intensity contrast between tissues complicates the differentiation of structures with shared attributes, a common challenge in soft tissue imaging \cite{shomal2024dynamic}. Discrepancies among imaging protocols, including the use of intravenous contrast agents, or sequence adjustments like acquisition time after administering contrast agents, significantly impact image characteristics. In addition, longitudinal inconsistencies caused by technological advancements (e.g., transitioning from 1.5T to 3T MRI systems) further exacerbate these issues \cite{lee2021benign}.

Medical image data also presents challenges related to its dimensionality and associated metadata. Variability in voxel spacing, along with heterogeneity across imaging modalities and protocols, leads to notable differences in resolution and comparability between different studies 
\cite{joutard2024hyperspace}. Anisotropic voxels, which arise from discrepancies in slice thickness and inter-slice gaps in CT and MRI, increase the complexity of data analysis \cite{liu2018influence}. The presence of anisotropic voxels can distort the representation of anatomical structures, which is especially problematic when trying to detect subtle changes, such as in brain tumor studies or vascular pathologies. Furthermore, inconsistencies in file formats, such as the original DICOM and its conversion into 
alternative simpler formats, hinder data sharing and interoperability, potentially resulting in critical misinterpretations of contextual information stored in metadata headers \cite{cabeen2017comparative, cobo2023enhancing}. Medical data interoperability is crucial not only for accurate diagnosis but also for continuity of care over time. Inconsistencies in formats can create gaps in critical information stored in metadata, which could lead to incorrect diagnoses or delays in patient care.

DL and generative AI algorithms have shown potential to address the aforementioned issues. Super-resolution techniques enable the reconstruction of high-resolution images from low-resolution inputs, facilitating the detection of small lesions and intricate anatomical details, particularly in LDCT and MRI \cite{li2023medical}. Image enhancement and denoising methods, including recent algorithms that leverage autoencoders and GANs \cite{qiu2023medical}, improve tissue differentiation and overall image quality by mitigating contrast and noise limitations \cite{chen2023deep}. 
Moreover, multimodal fusion techniques integrate data from multiple imaging modalities, providing a more comprehensive representation of anatomical and pathological features \cite{huang2020review}. 
Hybrid methods that incorporate physics-based constraints into DL frameworks, known as physics-informed machine learning (PIML), further enhance image reconstruction by combining data-driven and theoretical approaches for improved performance \cite{chen2023deep}, as will be discussed in the next section. Furthermore, advanced convolutional neural networks are demonstrating great potential in the precise segmentation of 3D images, allowing for the identification of complex structures, such as blood vessels or tumors, with greater accuracy. These techniques are beginning to revolutionize the fields of functional MRI  and CT imaging, improving surgical planning and early disease detection.

\section{Physics-informed machine learning}
\label{sec:physics_informed}
Throughout the previous section we have given examples of physics inspired algorithms in different medical imaging modalities. These algorithms leverage fundamental laws of physics underlying medical images to enhance AI applications, bridging the gap between natural image and medical image computer vision. The study of physics informed medical imaging has the potential to enhance explainability, consistency, physical plausibility, robustness and generalization. Prior knowledge from medical images can act as a regularization mechanism to constrain the range of acceptable solutions \cite{raissi2019physics}. For example, physics can be leveraged in generative models in the form of constraints to avoid creating non-realistic images, such as modeling echo attenuation \cite{dominguez2024diffusion}, reflection and scattering in ultrasound imaging, or simulating T1 and T2 relaxation phenomena in MRI \cite{planchuelo2024optimisation}, among many others. Hence, research into PIML algorithms has experienced an exponential growth in the last years, and we expect it to continue evolving. In this section, we will focus on the main characteristics of these hybrid methods where laws of physics are combined with artificial intelligence to build more reliable algorithms. 

Physics-informed learning can be defined as the process of leveraging prior knowledge derived from observational, empirical, physical, or mathematical understanding of the world to enhance the performance of a learning algorithm \cite{karniadakis2021physics}. Existing approaches are grouped by their use of physical principles to modify input data (observation bias), training losses (learning bias), and  network architectures (inductive bias), as explained in recent general reviews \cite{banerjee2024physics, karniadakis2021physics}, and a targeted review focused on medical image analysis tasks \cite{banerjee2024pinns}. These approaches can be combined to enhance PIML systems, resulting in more sophisticated hybrid methods. In the following paragraphs we give a general overview of existing approaches, and conclude with the main challenges in this field in relation with medical imaging.
\subsection{Observational biases}
\label{sec:obs_biases}
Observational data reflecting the underlying physics of the system is the simplest method of introducing biases in AI algorithms \cite{karniadakis2021physics}. Deep neural networks (DNNs) are trained to capture the underlying physical processes through exposure to diverse data during training \cite{banerjee2024physics}. The challenge is collecting large enough amounts of data for the system to reinforce these biases and generate robust predictions. AI generated synthetic data can be used for data augmentation, mitigating the need for large amounts of data. However, generating synthetic realistic medical images poses its own difficulties \cite{koetzier2024generating}.

\subsection{Learning biases}

Learning biases are prior assumptions that implicitly embed prior knowledge in the learning algorithm through soft penalty constraints (regularization) added in the loss function \cite{banerjee2024physics}. Learning biases can be expressed as integral, differential or even fractional equations to promote convergence towards physical plausible solutions. However, the underlying laws of physics can only be approximately satisfied \cite{karniadakis2021physics}. 
Physics-informed neural networks (PINNs) incorporate the knowledge of the physics of the process in the form of partial differential equations (PDEs) that are embedded into the loss function using automatic differentiation to calculate differential operators \cite{banerjee2024physics}. More recently, Sef-adaptive PINNs (SA-PINNs) allow adaptive training of neural networks by applying trainable weights to each training point, which enable to focus on challenging regions of the solution space \cite{mcclenny2023self}.

\subsection{Inductive biases}
\label{sec:ind_biases}
Inductive biases can be seen as a hard generalization of learning biases, such that prior assumptions are directly forced into the architecture of the model through specific design interventions, ensuring that the resulting predictions inherently comply with a defined set of physical laws. Examples are the well-known convolutional neural networks (translational invariance), but also capsule networks (translational equivariance) \cite{juralewicz2021capsule}, transformers (permutation equivariance) \cite{xu2024permutation}, graph neural networks, kernel methods such as Gaussian processes, Hamiltonian and Lagrangian neural networks, Neural ODEs \cite{peng2022learning} and more general PINNs, utilizing kernels directly derived by the fundamental physical principles of the problem \cite{banerjee2024physics, karniadakis2021physics}.

\subsection{Challenges}
\label{sec:physics_informed_challenges}
Introducing physics constraints into AI models to build more trustworthy medical image applications requires finding an optimal balance between the complexity of physics-based constraints and data-driven approaches to better capture real world dynamics and enhance generalization \cite{banerjee2024physics}. Domain expertise is necessary for selecting the most suitable physics prior to be modeled in the algorithm, and the best approach to introduce it in the model should be carefully considered. Incorporating excessive constraints during training can lead to over-fitting and over-regularization, therefore, ablation studies and quantitative assessments are necessary to evaluate the influence of such constraints on the model's performance \cite{banerjee2024pinns}. Additionally, it is essential to consider the scalability of these models in real-world medical environments, as incorporating too many physical constraints could limit the model's adaptability across different clinical settings. Although PIML is a promising direction of research in medical imaging, there remain limitations to the explainability of the model, such as understanding how physical constraints interact with learned features \cite{banerjee2024physics}; managing uncertainty to avoid overconfident models, and dealing with incompleteness of knowledge due to the inherent difficulties of modeling all possible phenomena. 

\section{Conclusions, trends and takeaways}
\label{sec:concl}
This work aims to emphasize the need of physics expertise in AI developers working on medical imaging, and serve as a reference for consultation, both for beginners in the field and those who seek to broaden their knowledge. A comprehensive understanding of the underlying physical properties in medical images for AI applications enhances explainability, and provides more reliable deep learning architectural designs, particularly for applications in image generation and reconstruction. In clinical settings, AI models typically face OOD data, which challenges reliability and overconfidence of AI algorithms, and in practice hinders their translation into clinical practice. PIML has shown promising potential to integrate physics into AI algorithms, thereby creating more trustworthy AI applications, and bridging the gap between natural and medical image computer vision.

This work has placed significant focus on generative AI, that is reshaping the landscape of medical imaging. Both medical image reconstruction and synthetic image generation require a specific understanding of the acquisition process behind each medical imaging modality. Generative AI is a dynamic and evolving field, however, the fundamentals of physics are always necessary for a full comprehension of medical images, and cannot be overlooked when developing generative AI models. Yet, it is crucial to acknowledge that, while these models hold significant promise, its integration into clinical practice must be carefully managed to avoid over-fitting or generating unrealistic synthetic data that could potentially mislead clinicians. 
Vision foundation models incorporating data from different medical image modalities have emerged to improve medical image analysis tasks. These models hold the promise to combine different modalities (multimodal) with various data types (text, image, video) and scales (cell, tissue, organ, patient, population) to assist radiologists and surgeons in their work. We envision the future of AI in medical imaging as a synergy between clinicians, physicists and AI developers, where AI algorithms support the workflow of medical professionals. This synergy will enable AI to move from a tool of analysis to a true collaborative partner in clinical decision-making.

The clinical translation of AI systems in medical imaging relies on building trustworthy algorithms that capture the complexity of real world data. We argue that physics plays a fundamental role in embedding prior knowledge into AI algorithms, providing complementary information to current data-driven methods with the aim of enhancing the learning process.


\section*{Acknowledgments} 

M. C. was supported by the Ministry of Education of Spain (FPU grant, reference FPU21-04458).

\section*{Declaration of competing interest}

The authors declare that the research was conducted in the absence of any commercial or financial relationships that could be construed as a potential conflict of interest.

\section*{Ethics statement}

The authors declare that this manuscript represents entirely original works, and/or if work and/or works of others have been used, that this has been appropriately cited or quoted. This material has not been published in whole or in part elsewhere. The manuscript is not currently being considered for publication in another journal.

\bibliographystyle{plain}
\bibliography{references}  

\end{document}